\documentclass[letterpaper,twocolumn,10pt]{article}
\usepackage{usenix2019_v3}

\usepackage{placeins}
\usepackage{amsmath}
\usepackage{graphicx}
\usepackage{xcolor}
\hypersetup{colorlinks=true,allcolors=blue}
\usepackage{booktabs}
\usepackage{array}
\usepackage{url}
\usepackage{setspace}
\usepackage{subcaption}
\usepackage{dblfloatfix}   

\renewcommand{\em}{\it}

\newcommand{\ignore}[1]{}

\def\cfigure[#1,#2,#3]{
\begin{figure}
\vspace*{0mm}
\begin{center}

\includegraphics[width=3in]{#1} 
 
\vspace*{-3mm}\caption[]{#2
} \label{#3}
 
\vspace*{-5mm}
\end{center}
\end{figure}}

\def\cfigurefour[#1,#2,#3]{
\begin{figure}
\vspace*{0mm}
\begin{center}

\includegraphics[width=4in]{#1} 
 
\vspace*{-3mm}\caption[]{#2
} \label{#3}
 
\vspace*{-5mm}
\end{center}
\end{figure}}

\def\cfiguretemp[#1,#2,#3]{
\begin{figure}
\vspace*{0mm}
\begin{center}

\includegraphics[width=3.5in]{#1} 
 
\vspace*{-3mm}\caption[]{#2
} \label{#3}
 
\vspace*{-5mm}
\end{center}
\vspace*{-2mm}
\end{figure}}

\def\wfigure[#1,#2,#3]{
\begin{figure*}
\vspace*{0mm}
\begin{center}
 \includegraphics[width=\textwidth]{#1} 
 \vspace*{-3mm}\caption[]{#2
} \label{#3}
 
\end{center}
\end{figure*}}

\def\threefigure[#1,#2,#3,#4,#5]{
\begin{figure*}
\vspace*{0mm}
\begin{center}

\begin{tabular}{ccc}
\includegraphics[width=2in]{#1} & \includegraphics[width=2in]{#2} &  \includegraphics[width=2in]{#3} \\
(a) & (b) & (c) \\
\end{tabular}

\vspace*{-3mm}\caption[]{#4
} \label{#5}

\vspace*{-5mm}
\end{center}
\vspace*{-2mm}
\end{figure*}}

\def\dcfigure[#1,#2,#3,#4,#5,#6]{
{
\begin{figure*}
\begin{center}
\begin{minipage}[c]{\columnwidth}{
\includegraphics[width=\columnwidth]{#1} 
\vspace*{0mm}\caption[]{#2} \label{#3} \
}\end{minipage}\hspace*{\columnsep}\
\begin{minipage}[c]{\columnwidth}{
\includegraphics[width=\columnwidth]{#4} 
\vspace*{0mm}\caption[]{#5}\label{#6} \
}\end{minipage}
\end{center}
\end{figure*}
}
}

\def\tableByTable[#1,#2,#3,#4,#5,#6]{
{
\begin{table*}
\begin{center}
\begin{minipage}[c]{3in}{
\centering
{#1}
\vspace*{0mm}\tabcaption[]{#2}\label{#3} \
}\end{minipage}\hspace*{\columnsep}\
\begin{minipage}[c]{3in}{
\centering
{#4}
\vspace*{0mm}\tabcaption[]{#5}\label{#6} \
}\end{minipage}
\end{center}
\end{table*}
}
}

\def\figureByTable[#1,#2,#3,#4,#5,#6]{
{
\begin{figure*}
\begin{center}
\begin{minipage}[c]{3in}{
\centering
\includegraphics[width=\textwidth]{#1}
\vspace*{0mm}\figcaption[]{#2} \label{#3} \
}\end{minipage}\hspace*{\columnsep}\
\begin{minipage}[c]{3.3in}{
\centering
{#4}
\vspace*{0mm}\tabcaption[]{#5}\label{#6} \
}\end{minipage}
\end{center}
\end{figure*}
}
}

\def\tableByFigure[#1,#2,#3,#4,#5,#6]{
{
\begin{figure*}
\begin{center}
\begin{minipage}[c]{4.3in}{
\centering
{#1}
\vspace*{0mm}\tabcaption[]{#2} \label{#3} \
}\end{minipage}\hspace*{\columnsep}\
\begin{minipage}[c]{2.2in}{
\centering
\includegraphics[width=\textwidth]{#4}
\vspace*{-0.35in}\caption[]{#5}\label{#6} \
}\end{minipage}
\end{center}
\end{figure*}
}
}

\def\doublecfigure[#1,#2,#3,#4]{
{
\begin{figure}
\begin{center}
\begin{minipage}[c]{1.5in}{
\begin{center}
\includegraphics[width=1.5in]{#1}
\end{center}
}\end{minipage}\hspace*{1em}\
\begin{minipage}[c]{1.5in}{
\begin{center}
\includegraphics[width=1.5in]{#2}
\end{center}
}\end{minipage}
\vspace*{0mm}\caption[]{#3} \label{#4} \
\end{center}
\end{figure}
}
}

\def\qcfigure[#1,#2,#3,#4,#5,#6]{
{
\begin{figure*}
\vspace*{0.2in}\
\begin{center}
\begin{minipage}[c]{3in}{
\includegraphics[width=3in]{#1} 
\vspace*{-3mm}
}
\end{minipage}\hspace*{0.5in}\
\begin{minipage}[c]{3in}{
\includegraphics[width=3in]{#2} 
\vspace*{-3mm}
}\end{minipage}

\begin{minipage}[c]{3in}{
\includegraphics[width=3in]{#3} 
\vspace*{-3mm}
}
\end{minipage}\hspace*{0.5in}\
\begin{minipage}[c]{3in}{
\includegraphics[width=3in]{#4} 
\vspace*{-3mm}
}\end{minipage}
\end{center}
\caption[]{#5}\label{#6}
\end{figure*}
}
}

\def\twfigure[#1,#2,#3,#4,#5]{
{
\begin{figure*}
\vspace*{0.2in}\
\begin{center}
\begin{minipage}[c]{6.5in}{
\includegraphics[width=6.5in]{#1} 
\vspace*{-3mm}
}
\end{minipage}

\begin{minipage}[c]{6.5in}{
\includegraphics[width=6.5in]{#2} 
\vspace*{-3mm}
}\end{minipage}

\begin{minipage}[c]{6.5in}{
\includegraphics[width=6.5in]{#3} 
\vspace*{-3mm}
}
\end{minipage}
\end{center}
\caption[]{#4}\label{#5}
\end{figure*}
}
}

\def\dwfigure[#1,#2,#3,#4]{
{
\begin{figure*}
\vspace*{0.2in}\
\begin{center}
\begin{minipage}[c]{6.5in}{
\includegraphics[width=6.5in]{#1} 
\vspace*{-3mm}
}
\end{minipage}

\begin{minipage}[c]{6.5in}{
\includegraphics[width=6.5in]{#2} 
\vspace*{-3mm}
}\end{minipage}

\end{center}
\caption[]{#3}\label{#4}
\end{figure*}
}
}

\def\dssfigure[#1,#2,#3,#4,#5,#6]{
{
\begin{figure*}
\vspace*{0.2in}\
\begin{center}
\begin{minipage}[c]{4in}{
\includegraphics[width=4in]{#1}
\vspace*{-3mm}\caption[]{#2} \label{#3} \
}\end{minipage}\hspace*{0.5in}\
\begin{minipage}[c]{2in}{
\includegraphics[width=2in]{#4}
\vspace*{-3mm}\caption[]{#5}\label{#6} \
}\end{minipage}
\end{center}
\vspace*{-0.4in}\
\end{figure*}
}
}

\def\dsfigure[#1,#2,#3,#4,#5,#6]{
{
\begin{figure*}
\vspace*{0.2in}\
\begin{center}
\begin{minipage}[c]{3in}{
\includegraphics[width=3in]{#1}
\vspace*{-3mm}\caption[]{#2} \label{#3} \
}\end{minipage}\hspace*{0.5in}\
\begin{minipage}[c]{3in}{
\hspace*{0.5in}\
\includegraphics[height=3in]{#4}
\vspace*{-3mm}\caption[]{#5}\label{#6} \
}\end{minipage}
\end{center}
\vspace*{-0.4in}\
\end{figure*}
}
}

\def\dsyfigure[#1,#2,#3,#4,#5,#6]{
{
\begin{figure*}
\vspace*{0.2in}\
\begin{center}
\begin{minipage}[c]{2.5in}{
\includegraphics[height=2.5in]{#1}
\vspace*{-3mm}\caption[]{#2} \label{#3} \
}\end{minipage}\hspace*{0.5in}\
\begin{minipage}[c]{2.5in}{
\includegraphics[height=2.5in]{#4}
\vspace*{-3mm}\caption[]{#5}\label{#6} \
}\end{minipage}
\end{center}
\vspace*{-0.4in}\
\end{figure*}
}
}

\def\dyfigure[#1,#2,#3,#4,#5,#6]{
{
\begin{figure*}
\vspace*{0.2in}\
\begin{center}
\begin{minipage}[c]{3in}{
\includegraphics[height=3in]{#1} 
\vspace*{-3mm}\caption[]{#2} \label{#3} \
}\end{minipage}\hspace*{0.5in}\
\begin{minipage}[c]{3in}{
\includegraphics[height=3in]{#4} 
\vspace*{-3mm}\caption[]{#5}\label{#6} \
}\end{minipage}
\end{center}
\vspace*{-0.4in}\
\end{figure*}
}
}

\def\dyoldfigure[#1,#2,#3,#4,#5,#6]{
{
\begin{figure*}
\vspace*{0.2in}\
\begin{center}
\begin{minipage}[c]{3in}{
\epsfysize=2.0in\
\hspace{0.5in}\
\epsfbox{#1}
\vspace*{-3mm}\caption[]{#2} \label{#3} \
}\end{minipage}\hspace*{0.25in}\
\begin{minipage}[c]{3in}{
\epsfysize=2.0in\
\hspace{0.5in}\
\epsfbox{#4}
\vspace*{-3mm}\caption[]{#5}\label{#6} \
}\end{minipage}
\end{center}
\vspace*{-0.4in}\
\end{figure*}
}
}

\def\cfiguredouble[#1,#2,#3,#4]{
\begin{figure}
\vspace*{0.2in}\
\begin{center}
\begin{minipage}[c]{1.5in}{
\epsfxsize=1.5in\
\epsfbox{#1}
}\end{minipage}\hspace*{0.1in}\
\begin{minipage}[c]{1.5in}{
\epsfxsize=1.5in\
\vspace{0.1in}\epsfbox{#2}
}\end{minipage}\vspace*{-0.10in} \caption[]{#3}\label{#4}
\end{center}
\vspace*{-0.4in}\
\end{figure}
}

\def\wpfigure[#1,#2,#3,#4]{
\begin{figure*}
\vspace*{4mm}
\begin{center}

\includegraphics[width=#4]{#1} 

\vspace*{-3mm}\caption[]{#2
} \label{#3}

\vspace*{-5mm}
\end{center}
\end{figure*}}

\def\wprfigure[#1,#2,#3,#4,#5]{
\begin{figure*}
\vspace*{4mm}
\begin{center}

\includegraphics[width=#4, angle=#5]{#1} 

\vspace*{-3mm}\caption[]{#2
} \label{#3}

\vspace*{-5mm}
\end{center}
\end{figure*}}

\def\DoubleFigureWSlide[#1,#2,#3,#4,#5,#6,#7,#8,#9]{
\begin{figure*}
\vspace*{#9}
\begin{center}
\begin{minipage}{#4}
\includegraphics[width=#4]{#1}
\vspace*{-3mm}\caption{#2
}\label{#3}
\end{minipage}
\hspace{2em}
\begin{minipage}{#8}
\includegraphics[width=#8]{#5}
\vspace*{-3mm}\caption{#6
}\label{#7}
\end{minipage}
\vspace*{-5mm}
\end{center}
\end{figure*}
}

\def\DoubleFigureW[#1,#2,#3,#4,#5,#6,#7,#8]{
\begin{figure*}
\vspace*{0in}
\begin{center}
\begin{minipage}{#4}
\includegraphics[width=#4]{#1}
\vspace*{-3mm}\caption{#2
}\label{#3}
\end{minipage}
\hspace{2em}
\begin{minipage}{#8}
\includegraphics[width=#8]{#5}
\vspace*{-3mm}\caption{#6
}\label{#7}
\end{minipage}
\vspace*{-5mm}
\end{center}
\end{figure*}
}

\def\DoubleFigureWHack[#1,#2,#3,#4,#5,#6,#7,#8]{
\begin{figure*}
\vspace*{0in}
\begin{center}
\begin{minipage}{3in}
\includegraphics[width=#4]{#1}
\vspace*{-3mm}\caption{#2
}\label{#3}
\end{minipage}
\hspace{2em}
\begin{minipage}{3in}
\includegraphics[width=#8]{#5}
\vspace*{-3mm}\caption{#6
}\label{#7}
\end{minipage}
\vspace*{-5mm}
\end{center}
\end{figure*}
}

\def\ddcfigure[#1,#2,#3,#4]{
\begin{figure*}
\vspace*{0.2in}\
\begin{center}
\begin{minipage}[c]{\columnwidth}{
\includegraphics[width=\columnwidth]{#1} 
}\end{minipage}\hspace{0.5in}\
\begin{minipage}[c]{\columnwidth}{
\includegraphics[width=\columnwidth]{#2} 
}\end{minipage} \caption[]{#3}\label{#4}
\end{center}
\end{figure*}
}

\def\ddcfigureSlide[#1,#2,#3,#4,#5]{
\begin{figure*}
\vspace*{#5}\
\begin{center}
\begin{minipage}[c]{3in}{
\includegraphics[height=3in]{#1} 
}\end{minipage}\hspace{0.5in}\
\begin{minipage}[c]{3in}{
\includegraphics[height=3in]{#2} 
}\end{minipage}\vspace*{-0.10in} \caption[]{#3}\label{#4}
\end{center}
\vspace*{-0.4in}\
\end{figure*}
}

\def\cxfigure[#1,#2,#3]{
\begin{figure}
\vspace*{4mm}
\begin{center}
 
\epsfxsize=2.5in\
\epsfbox{#1}\
 
\vspace*{-0.10in}\caption[]{#2
} \label{#3}
 
\vspace*{-5mm}
\end{center}
\vspace*{-2mm}
\end{figure}}

\newcommand{\beforecaption}{\vspace{-.15cm}\begin{spacing}{0.85}}
\newcommand{\aftercaption}{\vspace{-.45cm}\end{spacing}}

\newcommand{\eg}{\textit{e.g.}}

\newcommand{\sys}{TClone}

\usepackage{tikz}

\begin{document}
\raggedbottom
\date{}

\title{\Large \bf \sys: Low-Latency Forking of Live GUI Environments for Computer-Use Agents}

\author{
{\rm Yutong Huang}\\
University of California, San Diego
\and
{\rm Vikranth Srivatsa}\\
University of California, San Diego
\and
{\rm Alex Asch}\\
University of California, San Diego
\and
{\rm Hansin Tushar Patwa}\\
University of California, San Diego
\and
{\rm Yiying Zhang}\\
University of California, San Diego and GenseeAI
} 

\maketitle

\begin{abstract}
Computer-use agents increasingly operate inside live personal workspaces, where their actions can modify files, applications, GUI state, credentials, and authenticated sessions. This creates a tension between safety and quality: agents need isolation and rollback to avoid damaging user state, but also need fast branching to support speculative execution and parallel search. Existing VMs, containers, and checkpoint/restore systems can isolate or recover workloads, but they do not provide low-latency versioning of a full interactive workspace.

We present \sys{}, a forkable personal workspace system for computer-use agents. \sys{} enables a live GUI workspace to be snapshotted, forked into isolated branches, rolled back, and selectively committed or merged. Its design separates fast branch creation from durable checkpointing, using sibling containers, copy-on-write memory sharing, filesystem versioning, GUI-local execution, and asynchronous checkpointing. In our end-to-end agent-loop measurement, \sys{} reduces total task latency by 1.9$\times$ and 1.5$\times$ over KVM and CRIU. By making workspace versioning a first-class systems primitive, \sys{} supports safer and higher-quality agent execution over real personal computing environments.
\end{abstract}

\section{Introduction}
\label{sec:intro}

\textit{Computer-use agents} (CUAs) are beginning to operate inside the same interactive environments that people use every day. They browse the web, manipulate graphical applications, edit documents, run shell commands, use developer tools, and complete multi-step workflows over desktop-like workspaces~\cite{claudecomputeruse,claudecowork,openaicodexcloud,openclaw,hermesagent,osworld,agents}. Unlike traditional chat-based or API-driven AI assistants, these agents execute over ambient personal state: open windows, browser sessions, local files, credentials, application caches, terminal state, and in-memory program state. The personal workspace is therefore becoming an execution substrate for agentic computing.

This shift first creates a safety problem. CUAs are useful precisely because they can take real actions in real workspaces, but those actions can also cause persistent damage. A mistaken or malicious (\eg, prompt-injected) CUA operation can overwrite a document, delete a directory, change system settings, install or remove packages, or expose user credentials. The resulting state can be spread across files, application memory, GUI state, browser profiles, and local services. Safe CUA execution therefore needs a system-level \textit{undo} boundary around the workspace.

Today's CUA execution environments force users into unsatisfactory choices. Cloud-based workspaces leave the user's own PC intact, but the user needs to move their data and application context to a provider-controlled environment. A separate local machine or dedicated PC can contain the damage from a bad agent action, but it is expensive, inconvenient, and disconnected from the user's real working state. Running the agent directly on the user's day-to-day PC preserves context, but risks CUAs damaging critical user data and system settings. None of these options provides both personal-workspace fidelity and proper recovery from agent mistakes.

CUAs also raise a quality problem. Many tasks are long-horizon and exploratory: an agent may need to choose among several plans, recover from failed intermediate steps, or compare alternative executions before deciding which result to keep. Recent GUI-agent work shows that test-time tree search can substantially improve CUA success by exploring and evaluating alternative action trajectories~\cite{agentalpha}. However, to make this work for CUAs in practice, each trajectory needs its own coherent live workspace state, which is infeasible with today's PC environments.



CUAs' safety and quality problems both point to a missing link: a \emph{forkable, versioned workspace system}. At any point in an agent trajectory, the runtime should be able to turn the current personal workspace into a versioned state, create one or more isolated branches from it, and let those branches execute independently. Each branch should preserve the illusion of a complete PC environment, including processes, memory, GUI state, file-system state, and other I/O state, while making subsequent writes private to that branch. Failed branches should be cheap to discard or roll back to an older version. Successful branches should be promotable or mergeable under policy. In other words, CUAs need the equivalent of \texttt{fork()} and version control for an entire live interactive workspace, not merely for a single process or file tree.

This leads to the central question of this paper:

\begin{quote}
\itshape
Can we fork a live personal workspace cheaply enough for runtime versioned CUA execution?
\end{quote}

Existing virtualization and snapshotting systems provide useful pieces, but not this primitive. Virtual machines can snapshot rich machine state, but cloning a full VM is too heavyweight for frequent online branching. Containers are lightweight and can be checkpointed by tools like CRIU and Docker checkpointing~\cite{criu,dockercheckpoint}. However, these checkpointing tools are designed for container migration and thus adopt a synchronous checkpoint-and-restore process. 
Nor do they support complex live GUI session checkpointing.
Overall, existing solutions are all designed mainly for migration and backups, but CUA execution requires routine runtime versioning and forking.

In this paper, we present \textbf{\textit{\sys}}, a versioned personal workspace system for computer-use agents. \sys{} makes a live GUI workspace forkable: an agent runtime can create isolated branches from the current workspace, run agent trajectories inside those branches, discard failed branches, roll back to recorded versions, and promote a selected branch. The resulting abstraction supports both safety-oriented rollback and the common speculative pattern of \emph{fork, explore}, and \emph{select}.

The key technical challenge is making whole-workspace fork fast. A naive implementation would checkpoint the source workspace, serialize its memory and filesystem state, and restore a second copy before the agent can run. That approach is acceptable for migration or failure recovery, but not for online CUA execution.
We make a key observation that speculative workspace branches are initially almost identical to the state from which they are created. At fork time, applications, memory pages, GUI state, file contents, and cached data are largely shared. During execution, many branches diverge only modestly, and failed trials are often discarded before touching much state. 


Based on this observation, we propose to separate \emph{fast branch creation} from \emph{durable checkpointing}. The fast path reconstructs the workspace as a sibling container with independent namespaces, recreates the original process tree, installs copy-on-write mappings for anonymous memory across corresponding processes, creates a versioned filesystem view with shared read-mostly state, and runs the GUI stack inside the workspace container so display state belongs to the branch. The slow path captures durable state asynchronously, including dumping memory state to storage. Because of this separation, \sys{} can perform parallel container forks without any serialization point.

We implemented \sys{} by modifying the Linux kernel and the CRIU framework. We evaluate \sys{} on two CUA setups: AgentLoop running the GTA benchmark~\cite{gta,gta1}, and Agent S3~\cite{agents3} running OSWorld~\cite{osworld}, with a total of more than 600 computer-use tasks. We compare against CRIU-based checkpoint/restore, KVM-based cloning, and conventional container/filesystem baselines. \sys{} significantly improves the container fork time, by up to 4.9x and 3.4x compared to KVM and CRIU. Even when considering slower LLM (GPT-5.5) calls, \sys{} still improvs end-to-end latency of CUA tasks by a factor of 1.9$\times$ and 1.5$\times$ over KVM and CRIU. These results show that forkable personal workspaces can provide practical rollback and speculative execution at interactive timescales.

In summary, this paper makes the following contributions:
\begin{itemize}
    \item The introduction of \emph{versioned personal workspaces}: an OS abstraction for safe and efficient execution of CUAs. 
    \item The design of a fast workspace-fork architecture that separates online branch creation from durable checkpointing.
    \item Key mechanisms in workspace fork, including process, memory, file system, network, GUI, and security support.
    \item Comprehensive ealuation of CUA workspace behavior.
\end{itemize}

We will release \sys{} upon the acceptance of the paper.

\section{Background and Motivation}

This section gives background in CUA and CUA execution environments, and then motivates why current systems are insufficient.

\begin{figure*}[t]
    \centering
    \includegraphics[width=1.0\textwidth]{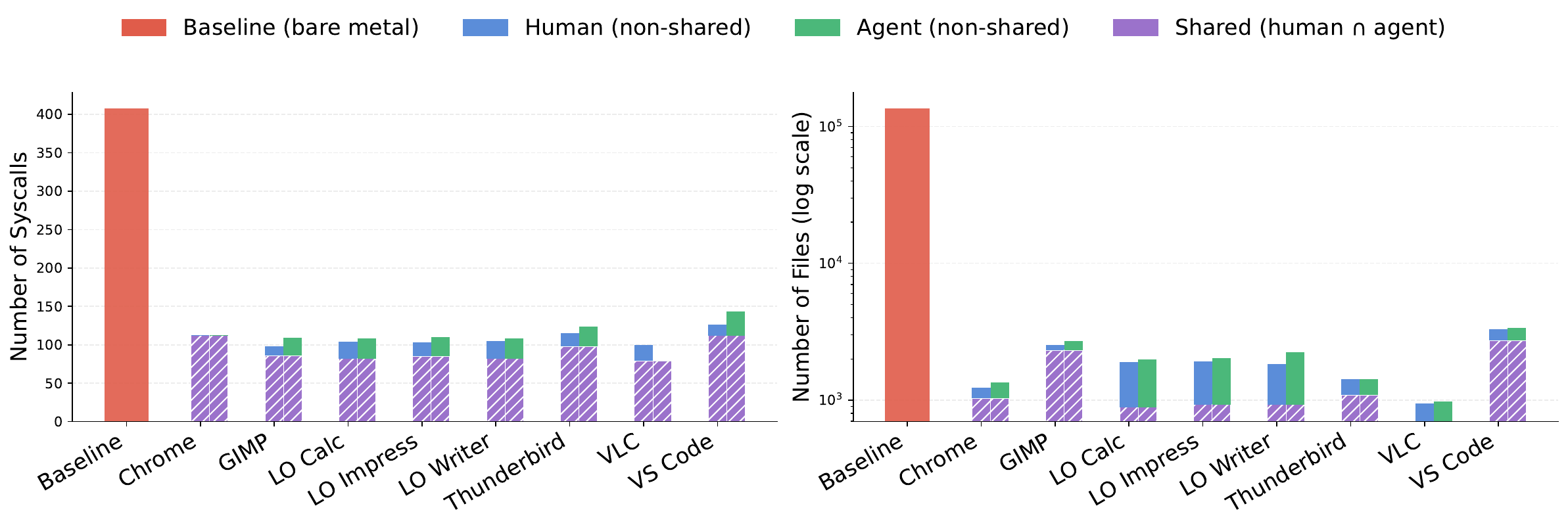}
    \caption{
    \textbf{OSWorld Task System Call and File Accesses.} Baseline: entire syscall and file system sets. Agent: running Agent S3. Human: human operating the same task. 
    }
    \label{fig:motivation-attack-surface}
\end{figure*}

\subsection{Computer-Use Agents}

Recent advances in foundation models have enabled \emph{computer-use agents}: agents that operate computers through human-facing interfaces such as GUIs, browsers, terminals, filesystems, and desktop applications. These agents observe screenshots, accessibility trees, browser DOMs, terminal output, or application state, and produce actions such as mouse clicks, keystrokes, shell commands, code edits, and tool invocations. This enables them to perform open-ended tasks across software environments rather than being limited to narrow APIs.

A growing set of systems demonstrates this paradigm. OpenClaw and Hermes Agents explore general-purpose desktop and personal-assistant workflows. Claude Computer Use and Claude Cowork allow agents to interact with applications, browsers, files, and development tools. OpenAI Codex supports autonomous software-engineering tasks through code editing and execution. Research systems such as GTA1 and Agent S3 study GUI-grounded agents that perceive screens, plan actions, and complete long-horizon computer tasks~\cite{openclaw,hermesagent,anthropiccomputeruse,anthropiccowork,openaicodex,gta1,agents3}.

Agent quality is increasingly tied to test-time computation. For language tasks, methods such as best-of-$N$ sampling, beam search, self-consistency, and tree search improve accuracy by exploring multiple candidate solutions before selecting one~\cite{selfconsistency,treeofthoughts,testtimescalingagents}. Recent CUA work makes the same point for GUI agents: Agent Alpha uses step-level tree search to explore, prune, and reuse partial trajectories, improving OSWorld success under comparable compute~\cite{agentalpha}. However, branching a computer-use agent is much harder than branching text generation: each branch may require its own browser state, GUI state, application memory, filesystem contents, terminal sessions, credentials, and network context. Conventional PCs are not designed to cheaply clone and run many independent desktop states in parallel.

\begin{figure*}[t]
\centering
\begin{subfigure}[t]{0.925\textwidth}
    \centering
    \includegraphics[width=\linewidth]{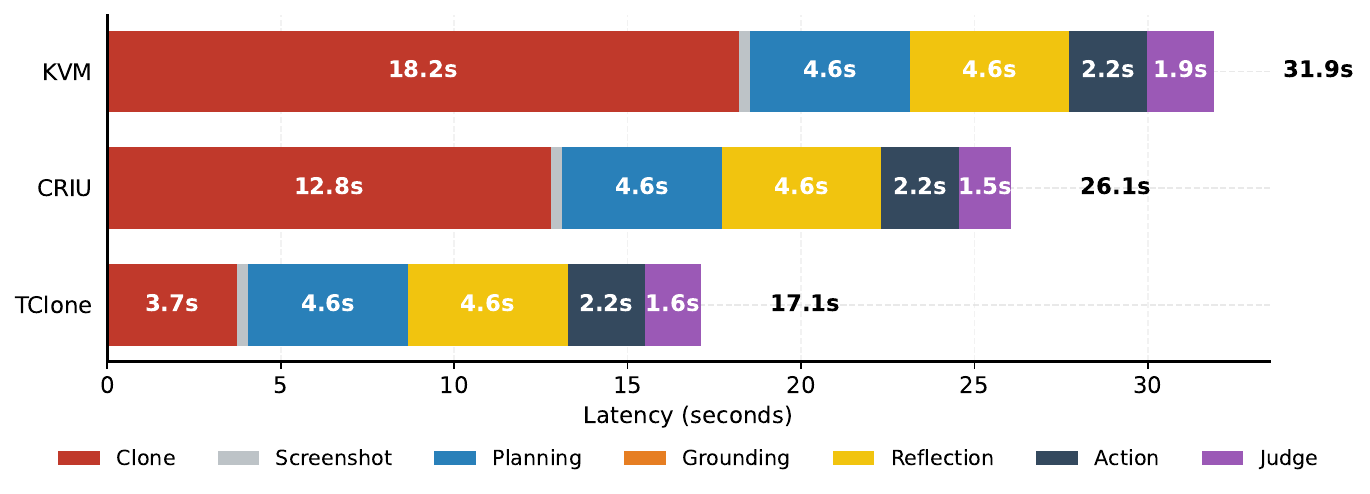}
    \caption{One branch point.}
    \vspace{0.1in}
    \label{fig:motivation-latency-breakdown}
\end{subfigure}
\hfill
\begin{subfigure}[t]{0.9\textwidth}
\centering
    \includegraphics[width=\linewidth]{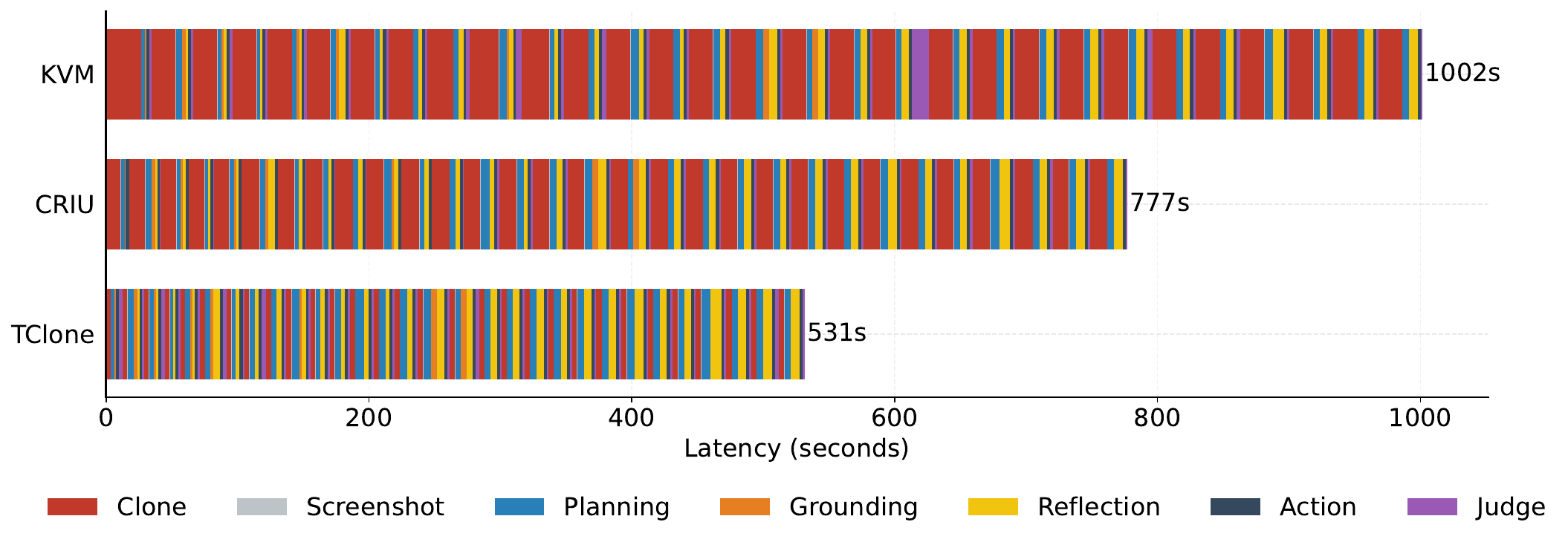} 
    \caption{Full agent execution.}
    \label{fig:motivation-timeline-breakdown} 
    \end{subfigure}
    \caption{ \textbf{CUA Execution Time Breakdown.} 
} 

\label{fig:motivation-branching-latency} 
\end{figure*}

\subsection{Safety and Security Implications}

Computer-use agents introduce safety and security risks because they can take real actions in real environments. Unlike chatbots that only produce text, these agents can modify files, execute commands, install packages, access private documents, interact with authenticated websites, and operate cloud or developer tools. A mistaken plan, hallucinated command, or adversarial instruction can therefore cause concrete damage.

A central problem is \emph{ambient authority}: the agent often inherits the privileges of the user or machine it controls. Desktop environments commonly contain browser cookies, SSH keys, API tokens, local repositories, personal files, and access to internal services. If the agent is compromised or confused, it may leak secrets, overwrite files, approve unintended actions, or run unsafe commands.

Figure~\ref{fig:motivation-attack-surface} quantifies this mismatch between authority and need on OSWorld~\cite{osworld} application tasks. The bare-metal baseline exposes more than 400 system calls and over \(10^5\) files. In contrast, both human and agent executions use much smaller application-specific subsets: across a variety of tasks such as browser and coding automation, the observed syscall footprint is roughly a quarter of the entire available syscalls, and the file footprint is orders of magnitude smaller than the full filesystem.
This suggests that blindly enabling CUAs to access the whole system is unnecessary and risky.


Parallel agent search can amplify these risks. If an agent runs many speculative branches, each branch may attempt external actions such as sending messages, submitting forms, pushing commits, or calling APIs. Existing desktops do not distinguish speculative actions from committed actions. Thus, safe parallel search requires isolation, rollback, external side-effect mediation, audit logs, and explicit commit mechanisms.

\subsection{Limitations of Existing Solutions}

Existing execution environments provide useful pieces, but none directly support stateful, interactive, and speculative computer-use agents. Virtual machines offer strong isolation and snapshot-based rollback, but cloning full guest state is expensive. Containers are lighter and widely used for command-line and coding agents, but they share the host kernel and often do not fully capture GUI state, browser profiles, desktop services, devices, credentials, and other personal-workspace state. MicroVMs improve the performance-isolation tradeoff for service workloads~\cite{firecracker}, but they are not optimized for rich desktop interaction.

Checkpointing systems add another building block. CRIU can preserve Linux process trees, including memory, file descriptors, namespaces, and related process state~\cite{criu}; VM and filesystem snapshots can preserve coarser-grained state. However, a useful CUA branch contains more than process memory or disk blocks: it includes browser tabs, cookies, GUI windows, display buffers, clipboard contents, filesystem mutations, terminal sessions, local services, network connections, and application caches. Existing mechanisms expose low-level snapshot and restore operations, but not a first-class abstraction for agent trajectories, rollback points, parallel branches, branch-specific policy, or controlled commit.


Rollback and speculation place workspace cloning on the critical path of an agent loop. Before a risky action, the runtime must preserve the current state; before parallel exploration, it must create one branch per candidate trajectory. If this takes seconds to tens of seconds, the agent spends more time preparing environments than doing useful work.

Figure~\ref{fig:motivation-branching-latency} shows the latency breakdown of an Agent S3~\cite{agents3} executing an OSWorld task~\cite{osworld}, including time spent on taking screenshots, model planning (what actions to take), model grounding (turning model output into actual action on PC), model self-reflection, actual computer-use actions, model judging (the effect of the action), and workspace cloning. We compare default KVM, CRIU, and \sys. All model calls are from GPT-5.5.

Figure~\ref{fig:motivation-latency-breakdown} shows the breakdown of a single step, and Figure~\ref{fig:motivation-timeline-breakdown} shows the end-to-end latency.
Contrary to common belief, LLM model execution accounts for less than half of the total execution time for default KVM and CRIU, because of their significant workspace cloning cost.
In contrast, \sys{}'s efficient container fork improves cloning latency by up to 4.9× and 3.4× over KVM and CRIU, and end-to-end latency by up to 1.9x to 1.5x, respectively.



\begin{figure*}[t]
    \centering
    \includegraphics[width=0.9\textwidth]{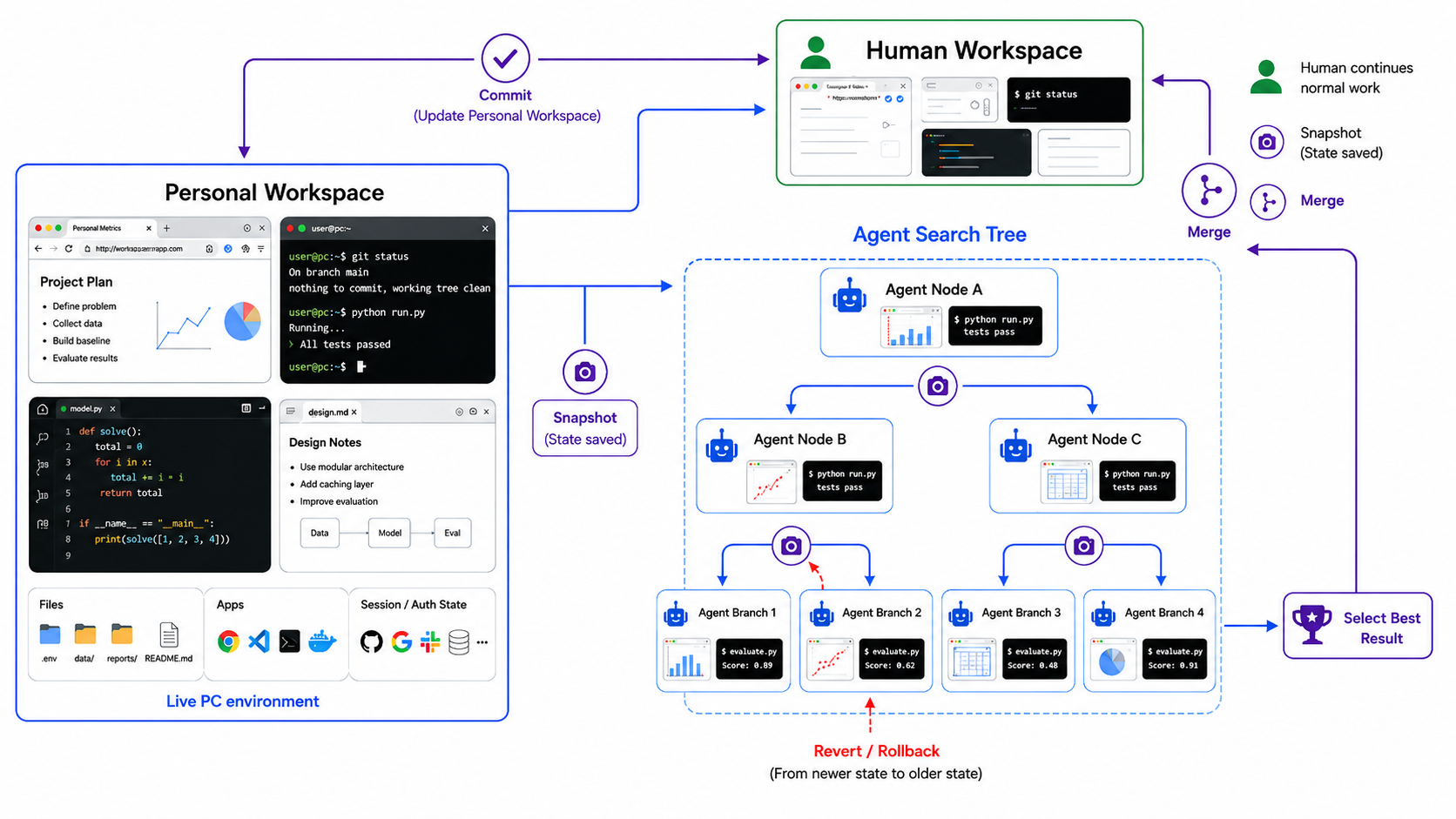}
    \caption{
     \textbf{Overview of \sys{} Personal Workspace Versioning.} 
    }
    \label{fig:sys-overview}
\end{figure*}

\section{\sys{} Overview}
\label{sec:overview}

\sys{} is an execution and orchestration system for computer-use agents in PC environments. \sys{} adopts a container-centered execution and versioning approach, where all workspace exeuction, regardless of from human or agents, is encapsulated in containers. Each container captures all application processes, memory state, file systems, GUI state, and other I/O states that human or agents would access. \sys{} handles the efficient snapshotting, cloning, rollback, and merging of containers.

Figure~\ref{fig:sys-overview} shows the lifecycle of a computer-use task (\eg, filling out an online form in browser). Before an agent task starts, \sys{} records a version of the current workspace container. It then forks an agent container from that version while leaving the original container available for the human. During the task, the agent runtime can fork additional containers for parallel exploration, roll a container back after a bad action, or discard failed branches. For example, an agent can try several GUI action sequences from the same browser state, keep the branch that reaches the intended result, and drop the rest. When the task completes, \sys{} selects one or more successful agent containers and merges or promotes their state into the user-visible workspace.

\sys{} exposes versions as branch points in an agent execution tree. Containers derived from the same version are isolated from one another: a failed agent branch cannot corrupt the human container or another candidate branch. At the same time, \sys{} heavily uses copy-on-write mechanisms for unchanged state, so branch creation does not require eagerly copying the full desktop state. Conceptually, \sys{} provides the semantics of process-level \texttt{fork()} at the granularity of a full {\em interactive GUI workspace}.

\sys{} supports three operations over this version tree that a CUA or a human can call. \emph{Fork} creates a runnable sibling container from an existing version. \emph{Rollback} returns one container to an earlier recorded version without affecting other containers. \emph{Commit} makes selected speculative state visible to the human workspace, either by promoting one chosen container or by merging selected state from multiple successful containers. These operations are policy boundaries as well as state-management mechanisms. \sys{} can attach different security profiles to agent containers, restrict external side effects, and require approval before sensitive state is committed. 
The rest of the paper describes how \sys{} makes this abstraction practical. 


\section{\sys\ Design}
\label{sec:design}


\begin{figure*}[t]
    \centering
    \includegraphics[width=0.8\textwidth]{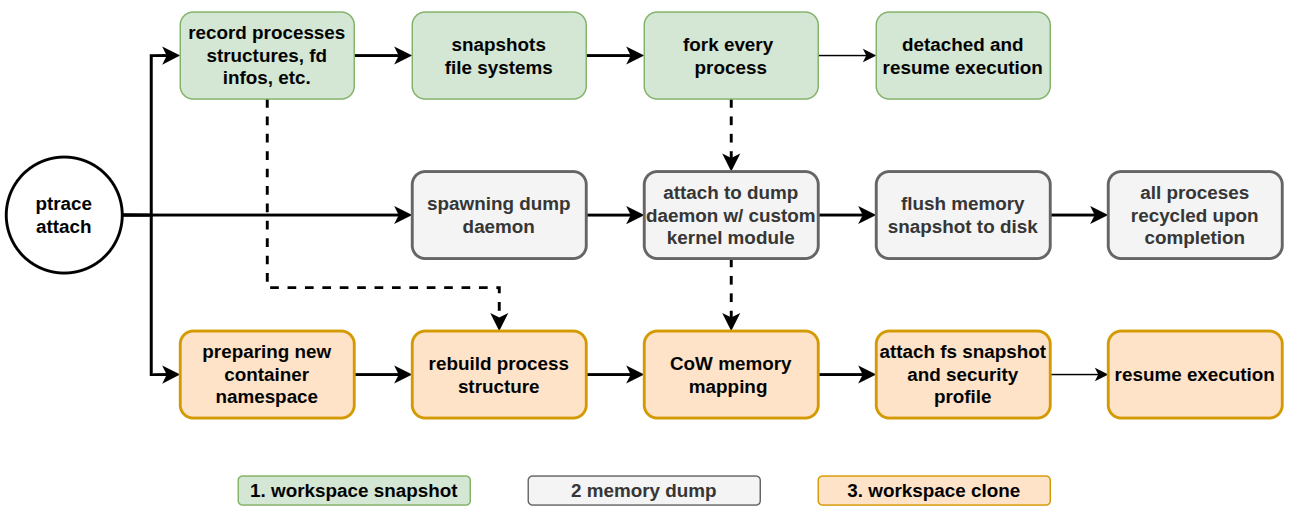}
    \caption{
     \textbf{\sys\ Workspace Fork Procedure.} \sys{} parallelize snapshot, clone, and memory state persistency
    }
    \label{fig:clone-procedure}
\end{figure*}

This section describes how \sys{} implements efficient whole-workspace forking and versioning.
A workspace branch contains several kinds of state that must be reconstructed or shared consistently: process state, anonymous memory,
file-backed memory, shared-memory regions, filesystem state, network state, and GUI state.
The design follows one architectural principle: \emph{online branch creation is separated from durable checkpointing}.
Branch creation reconstructs a running workspace using copy-on-write sharing and never copies page contents on the critical path;
durable serialization proceeds asynchronously and off that path. Figure~\ref{fig:clone-procedure} shows the overall workspace fork workflow.
Below we describe each subsystem and \sys{}'s security support.

\subsection{Process Subsystem}

\begin{figure*}[t]
    \centering
    \includegraphics[width=0.9\linewidth]{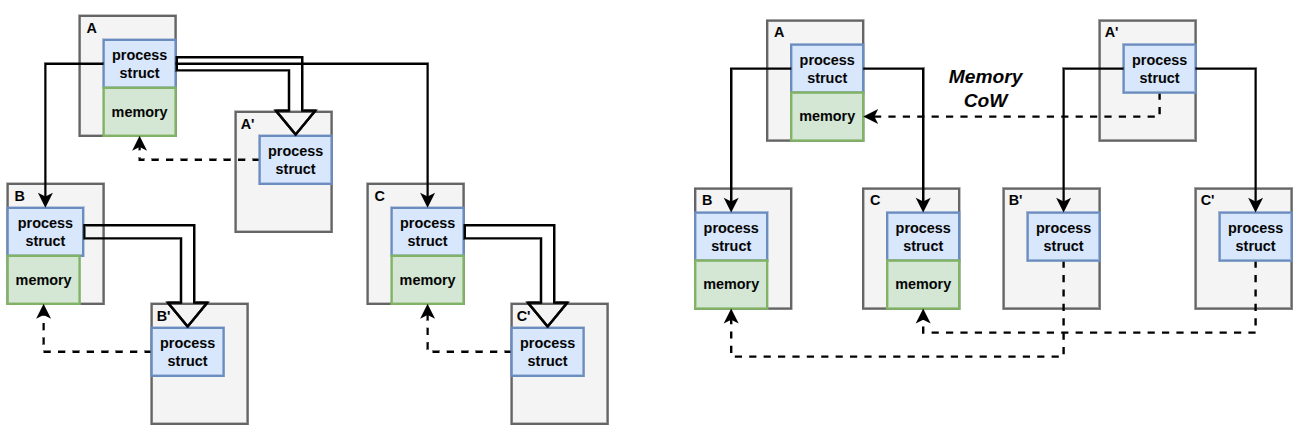}
    \caption{
    \textbf{Linux Process Fork vs. \sys\ Process-Tree Fork}. Left: native \texttt{fork()} duplicates one process as a child of the source process. Right: \sys{} creates a sibling workspace container and reconstructs the whole process tree inside an independent namespace.
    }
    \label{fig:process-fork}
\end{figure*}

To fork a workspace, \sys{} must first reproduce its process tree.
Using Linux \texttt{fork()} directly does not work: a forked process has the original process as its parent,
so individually forking processes $A$, $B$, and $C$ yields the inconsistent tree on the left of Figure~\ref{fig:process-fork},
in which $A'$, $B'$, and $C'$ are children of the source processes rather than of one another.
Cloning a workspace instead requires reproducing the \emph{original} tree topology---$A'$ as the parent of $B'$ and $C'$---inside an isolated container,
as shown on the right.

\sys{} reconstructs the tree rather than forking it in place.
The control plane briefly freezes the source container and, while it is stopped,
records the process hierarchy, thread structure, namespace membership,
open descriptors, signal and futex state, and register and TLS state needed to resume execution.
\sys{} then creates a destination container with its own PID, mount, network, IPC,
and UTS namespaces and replays the recorded topology:
a top-level restorer process recreates each task with its recorded namespace-local PID,
so the application-visible process tree is preserved
while host PIDs differ across branches and no PID conflicts arise between the human container,
the agent container, and other speculative branches.
Because the restorer executes in a fresh PID namespace, it cannot name source processes by integer PID;
\sys{} references each frozen process through a \texttt{pidfd} opened before the namespace transition,
which the memory subsystem uses to attach shared pages across the container boundary.

The freeze interval covers only metadata capture and the creation of the memory snapshot holders
described in \S\ref{sec:design-mem};
no page contents are copied while the source is stopped,
so the interval is independent of the workspace's memory and disk size.
Once metadata, snapshot holders, and the filesystem snapshot are consistent, the source processes resume;
durable serialization continues in the background.

\subsection{Memory Subsystem}
\label{sec:design-mem}


The primary opportunity in workspace forking comes from the execution pattern of CUA environments.
The environment is heavyweight, but an agent's speculative branch typically performs incremental actions
and touches only a small fraction of the address space before the branch is committed or discarded.
Eagerly duplicating the full memory footprint of each clone would impose prohibitive time and capacity overhead,
so \sys{} shares memory copy-on-write (CoW) across containers and diverges lazily.


\sys{} distinguishes three classes of memory.
\emph{Anonymous} memory is not backed by files.
\emph{File-backed} memory holds executable images, shared libraries, mapped files,
and application data, and is coordinated with the filesystem subsystem.
\emph{Shared} memory is used for inter-process communication and, in GUI workloads,
frequently carries rapidly mutating display buffers.
\sys{} applies CoW sharing primarily to anonymous memory;
constantly rewritten shared-memory regions such as framebuffers are treated as branch-local,
since sharing them would yield little reuse and inflate fault overhead.


Anonymous-memory sharing cannot be expressed with an ordinary \texttt{fork()},
because the two endpoints are corresponding processes in two independently isolated containers
rather than a parent and its immediate child.
\sys{} therefore decomposes the per-process address-space duplication that \texttt{fork()}
performs into a per virtual memory area (VMA) operation that the control plane can drive across the container boundary,
implemented as a small kernel module. A frozen source process is identified
through its \texttt{pidfd} so it is resolvable from the destination's namespace. The mechanism
duplicates the memory region of a source process into the corresponding destination process, links it into the source's
anonymous reverse-mapping (\texttt{anon\_vma}) chain, and installs the source's page-table
entries write-protected in the destination. No page contents are copied.
Subsequent sharing and divergence are handled entirely by the kernel's existing copy-on-write fault path:
the first write on either side allocates a private page for the writer through the standard
reverse-mapping machinery while the other side continues to reference the original page.
Because this is the same primitive \texttt{fork()} uses,
$N$ branches of one source are supported with no additional mechanism,
structurally equivalent to invoking \texttt{fork()} $N$ times against the same frozen parent.

Not every region can be shared by anonymous CoW without breaking isolation,
and preserving isolation is what makes a failed or malicious branch unable to perturb the human workspace.
File-backed regions are rebound to the branch's own filesystem view (\S\ref{sec:design-fs})
so that branch writes or incidental writes by the reconstruction path cannot reach the source's page cache;
intra-branch CoW is retained only for their already-private pages.
Regions backed by POSIX shared segments are reconstructed as independent segments
in the branch's namespace. With this per-class treatment,
after a branch is created a write on any side affects only that side's pages.

\begin{figure*}[t]
    \centering
    \includegraphics[width=\textwidth]{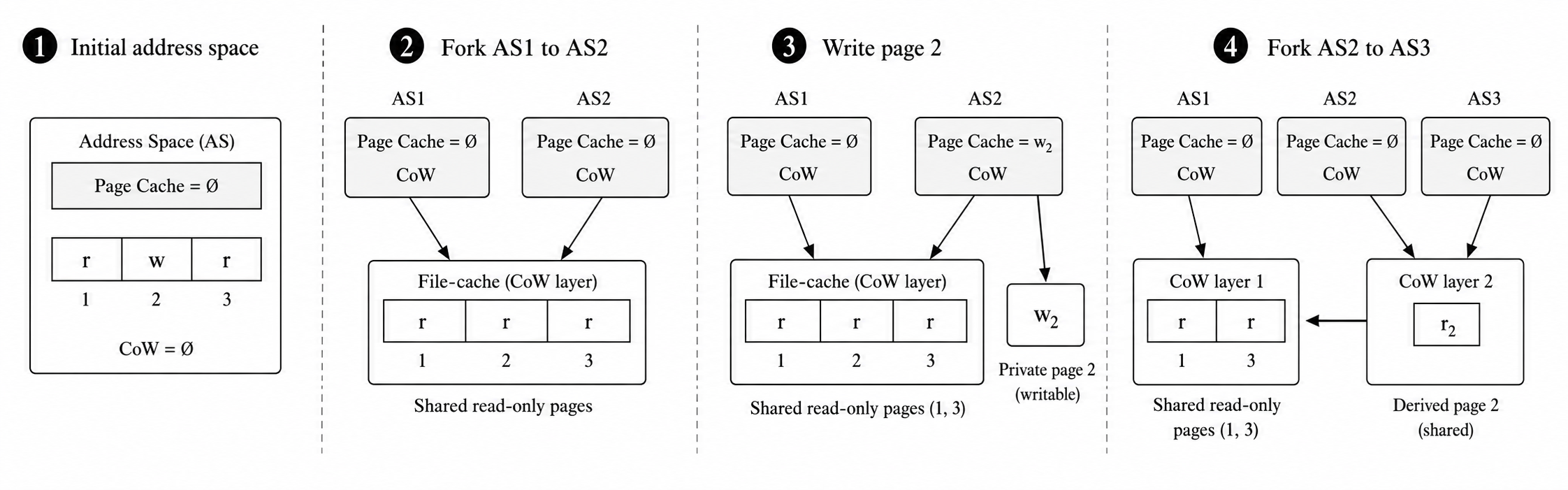}
    \caption{
     \textbf{Lazy CoW File-Cache Versioning.} \sys{} forking address space 2 and 3 from the original address space 1. 
    }
    \label{fig:file-cache-cow}
\end{figure*}

Rollback requires a stable image of a branch point,
but writing that image synchronously would reintroduce the latency that fast branching eliminates.
\sys{} separates logical snapshot creation from durable persistence.
While the source is frozen, \sys{} creates one snapshot holder per process using the kernel's standard copy-on-write
\texttt{fork()} (Figure~\ref{fig:clone-procedure});
each holder retains a consistent point-in-time view of its process's address space at no copy cost,
and the source may resume and mutate its pages without perturbing that view.
Holders are reparented to a background dump daemon that serializes them to storage and reclaims them on completion,
so neither the source nor any branch waits for durable serialization. 
Successive checkpoints of a long-running workspace chain through dirty-page tracking,
so each durable snapshot records only the pages modified since the previous one and
unchanged pages are referenced by the parent image.


\subsection{File System Subsystem}
\label{sec:design-fs}

Filesystem state has different semantics from conventional snapshots,
which are typically immutable recovery points while the active filesystem continues on a separate path.
In agent workloads many branches are live simultaneously and each may write; beam search,
best-of-$N$ execution, and rollback-oriented workflows
require multiple filesystem versions to coexist and diverge concurrently.
\sys{} therefore treats filesystem branching as an online versioning primitive rather than an
occasional backup operation, layering it on a snapshot-capable on-disk format that already performs block-level CoW.

Existing container filesystems fit this workload poorly.
Overlay-based filesystems expose one writable upper layer over read-only lower layers:
repeated branching produces deep lower-layer chains whose lookup latency degrades with depth,
a branch cannot be forked from another branch because each layer is a mountpoint,
and copy-up granularity defeats fine-grained progressive checkpoints.
A more fundamental problem is filesystem-independent and is not solved by block-level CoW:
the page cache is keyed per address space, so whenever a snapshotting filesystem gives a
snapshot its own inode, including block-CoW filesystems such as btrfs or ZFS, both branches
cache the same on-disk content twice in memory even though their extents are reflink-shared.
Block-level CoW thus deduplicates on disk but not in RAM.

\sys{} addresses the memory-duplication problem with a CoW-aware page cache,
structurally analogous to the kernel's existing cross-process sharing of identical anonymous pages
but applied to file-backed pages. The key difficulty is that filesystem CoW cannot mimic eager process CoW:
the page cache may hold dirty, not-yet-flushed pages, and a new branch may never touch most cached pages,
so eagerly write-protecting and duplicating cache state at fork time would add work and memory that most
branches never use. \sys{} instead uses lazy page-cache CoW. When a branch is created, \sys{} records a sealed,
read-only CoW layer shared by the related filesystem address spaces.
A branch first consults its private page cache; on a miss the kernel walks the shared layer chain before
falling back to storage, mapping any layer-resident page read-only into the branch.
A write to a shared cached page materializes a private copy in the writing branch and leaves the layer
untouched for the others; a further fork derives a new layer so siblings can share post-write state while
ancestors continue to observe the earlier version.

Two properties keep this correct under concurrent divergence.
First, layers are immutable once sealed, so a branch write always
copies out and a branch never observes another branch's modifications.
Second, because a source may modify a file between the on-disk snapshot point and the lazy page-cache fork,
a cached page is admitted into the shared layer only when the filesystem confirms its
underlying extent is still shared between source and branch;
pages whose extents have diverged are excluded, and the branch reads them from its own on-disk extent.
Figure~\ref{fig:file-cache-cow} illustrates the layered structure.
The result is a filesystem branch that behaves as a private writable version of the workspace
while sharing read-mostly state such as binaries, libraries, resources, and unchanged documents
across branches without eager copying.

\subsection{Network Subsystem}

Network state must distinguish communication internal to a workspace from communication with external services.
\emph{Internal} connections have both endpoints inside the same branch
(e.g. loopback connections between a browser and a local helper, an application and a local daemon,
or a GUI client and a local service)
and are part of the workspace state.
\sys{} captures their TCP state using kernel TCP-repair mechanisms and re-establishes them inside the
destination container, whose fresh network namespace gives each branch an independent loopback and
address space so reconstructed endpoints and ports do not collide across branches.

\emph{External} connections have different semantics:
if a workspace is cloned while connected to a remote server,
the server cannot safely treat both branches as the same endpoint,
and duplicating the connection would create ambiguity over ownership, ordering, replayed data, and side effects.
\sys{} therefore closes external connections during branch creation and
requires each branch to re-establish its own.
This makes the clone explicit to the remote endpoint and,
because external effects such as submitting a form or pushing data cannot generally be undone while local state can,
creates a natural boundary at which a speculative branch can run with restricted networking,
delayed external commit, or explicit approval before contacting remote services.



\subsection{GUI Subsystem}
\label{sec:design-gui}

Computer-use agents need a graphical environment that is both realistic and checkpointable. A design that runs applications in a container but depends on a host X11~\cite{X11} or Wayland~\cite{Wayland} server is problematic: the application state may be inside the container, while window state, input state, compositor state, clipboard contents, and display buffers remain on the host. Restoring only the container can therefore leave the GUI in an inconsistent state.

\sys\ avoids this problem by running the GUI stack inside each workspace container. Our implementation uses a Webtop~\cite{webtop} environment, leveraging Selkies~\cite{Selkies}, where a Wayland compositor and its GUI clients run together inside the container, and the desktop is exported over HTTPS. This makes the compositor part of the workspace state rather than an external host dependency. As a result, each forked branch has its own display session, cursor, windows, clipboard, and input stream, so human and agent branches cannot interfere with one another through shared GUI state.

This design also simplifies checkpointing. Since the compositor and applications are in the same container, \sys\ can reconstruct them together during fork and restore. Highly dynamic GUI buffers, such as framebuffer-like shared memory, are treated as branch-local state rather than aggressively shared with copy-on-write, while larger application memory and file-backed state can still benefit from sharing.

\subsection{Commit and Merge Support}

\sys{} treats commit as the point where speculative container state becomes user-visible state. The simplest policy promotes one selected agent container wholesale and discards the alternatives, matching best-of-\(N\) and beam-style execution where the runtime chooses a single successful trajectory.

When the human container and one or more agent containers have concurrent execution, \sys{} exposes merge as a controlled policy boundary. Persistent files are merged using existing file-level merge mechanisms, with policy or human approval for conflicts and sensitive paths. For volatile process state, \sys{} supports promotion by replacing a process address space with the selected agent container's address space. 

Note that \sys{} does not commit, merge, or rollback external side effects, such as sending emails or making online payments. \sys{} treats external effects as a policy boundary: speculative branches can run with restricted networking, delayed release of external actions, or explicit approval before effects are committed.

\subsection{Security Support}
\label{sec:design-security}

A forked agent branch should not automatically inherit the full authority of the user's workspace. Many agent tasks require only a narrow subset of the environment. For example, a browser or editor task rarely needs SSH keys, cloud credentials, private directories, shell history, or unrelated project files. \sys\ therefore assigns each agent branch a security profile at its fork time. To do so, we first capture a security profile for each application by recording syscalls, files, devices, and other resources accessed by the appliction when a human operates it. These are the allowlist of that application for agents to safely access. 

We use these pre-recorded human-based profiles to adaptively set security profiles for each container (based on what applications they access), without involving the agent in complex profile building, or allowing a potentially malicious agent to build a harmful profile.
\sys\ enforces security profiles with standard container-hardening mechanisms. Seccomp restricts the system-call surface, while SELinux controls access to labeled files, devices, sockets, and other resources. This moves protection out of prompts and agent behavior: a branch can be forked with only the resources needed for the task, preventing the agent from reading many secrets in the first place.


One caveat is that human traces are not complete because agents also invoke automation tools, such as Python scripts and \texttt{xdotool}, that humans do not use directly. In practice, the baseline hardened profile protects sensitive data for many application tasks, while task-specific profiles mainly add application-specific configuration paths. OS configuration and package-management tasks are harder: they require irregular and potentially privileged access across the filesystem, so \sys{} treats them as requiring separate policy rather than the default application profiles.

\section{Evaluation}

\label{sec:evaluation}

We implemented \sys{} by changing the Linux kernel and extened the CRIU system. We evaluate whether \sys{} makes versioned workspace execution practical for computer-use agents, including per-task end-to-end agent task execution latency, workspace clone latency, workspace memory footprint, and clone scalability.

\subsection{Experimental Setups and Baselines}

We evaluate two agent-benchmark setups. The first runs AgentLoop on GTA~\cite{gta}, using the OpenCompass GTA codebase and following the GTA1 test-time-scaling setup~\cite{gta1}. GTA contains 229 multimodal human-written tools with executable tool chains. The second runs Agent S3~\cite{agents3} on OSWorld~\cite{osworld}, which contains 369 open-ended desktop tasks across applications such as browsers, office tools, media players, and developer environments. We do not evaluate every agent on every benchmark; each result corresponds to one of these two setups.



We compare \sys{} against two baselines.
\textbf{CRIU} is stock container checkpoint/restore:
it preserves process state but creates a branch through a synchronous
checkpoint-and-restore path with serial restores;
the ablation in \S\ref{sec:eval-ablation}
decomposes exactly why this path is slow.
\textbf{KVM} is VM-based isolation and snapshotting:
it provides whole-machine isolation but pays the cost of cloning
full virtual-machine state per branch.
The microbenchmarks further illustrate the degradation in OverlayFS access latency inherent to conventional container filesystems in heavy snapshotting environments.
All experiments run on a desktop-class 13th-Gen Intel Core i5-13400 with 32\,GB DDR4.

\begin{figure*}[t]
    \centering
    \begin{subfigure}[t]{0.49\textwidth}
        \centering
        \includegraphics[width=\linewidth]{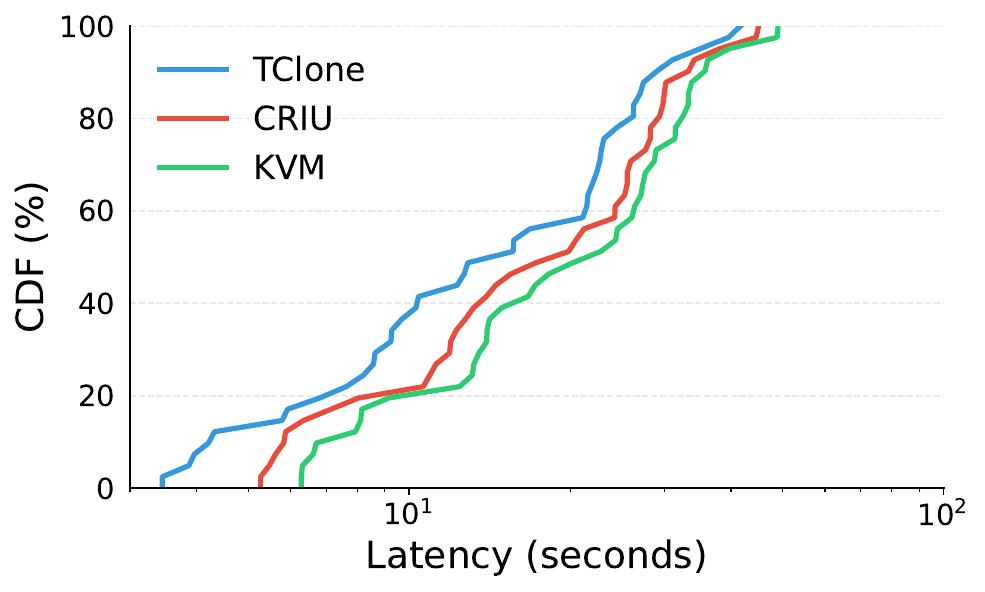}
        \caption{AgentLoop on GTA.}
        \label{fig:e2e-agentloop}
    \end{subfigure}
    \hfill
    \begin{subfigure}[t]{0.49\textwidth}
        \centering
        \includegraphics[width=\linewidth]{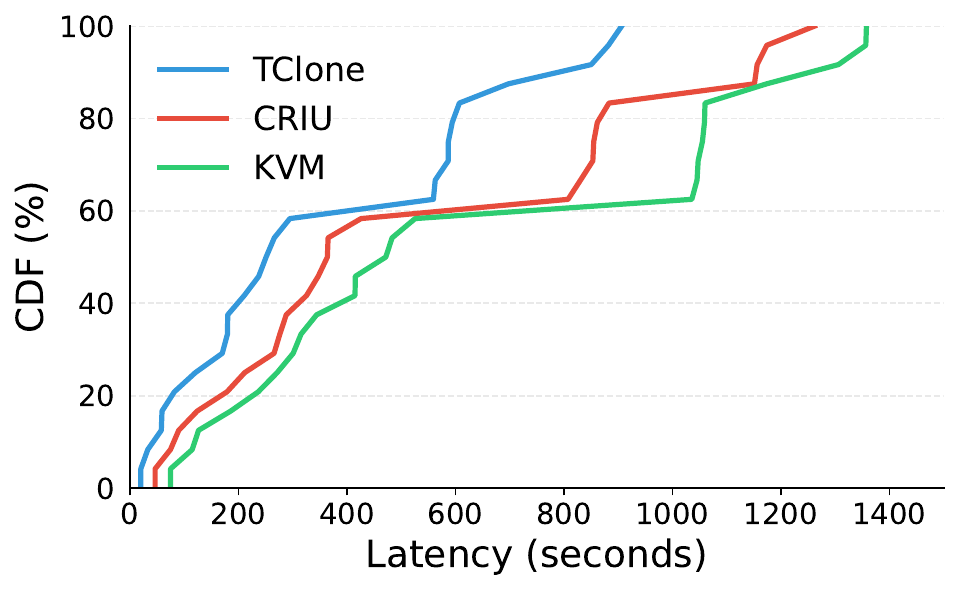}
        \caption{Agent S3 on OSWorld.}
        \label{fig:e2e-osworld}
    \end{subfigure}
    \caption{
     \textbf{CDF of end-to-end task latency.} Left: AgentLoop running the GTA benchmark. Right: Agent S3 running the OSWorld benchmark.
    }
    \label{fig:e2e-results}

\end{figure*}

\subsection{End-to-End Latency}

\begin{figure*}[t]
    \centering
    \begin{subfigure}[t]{0.49\textwidth}
        \centering
        \includegraphics[width=\linewidth]{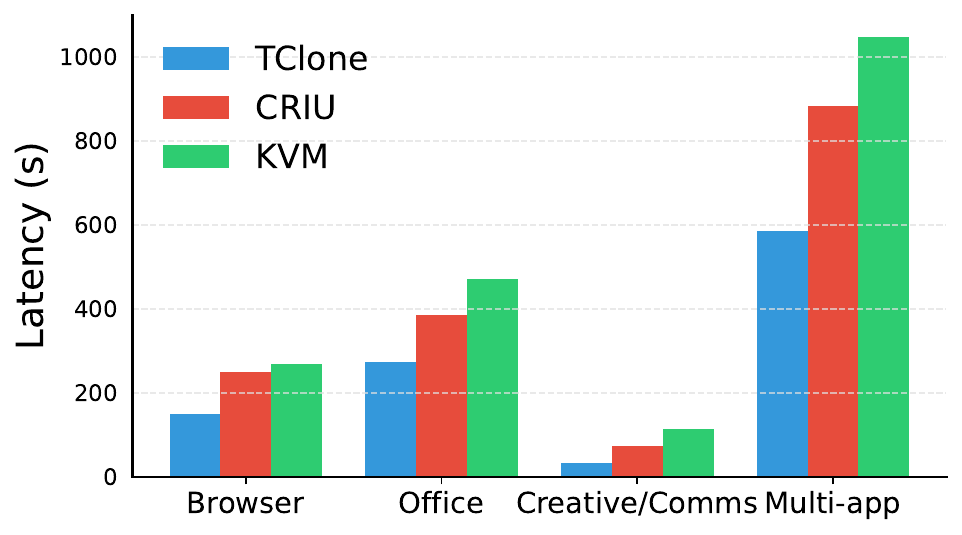}
        \caption{Latency by category.}
        \label{fig:osworld-category-latency}
    \end{subfigure}
    \hfill
    \begin{subfigure}[t]{0.49\textwidth}
        \centering
        \includegraphics[width=\linewidth]{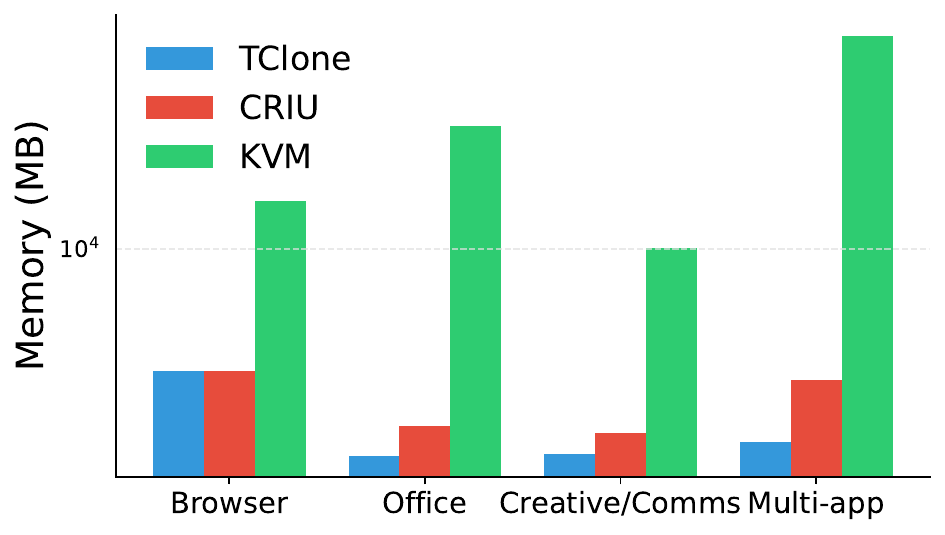}
        \caption{Memory footprint by category.}
        \label{fig:osworld-category-memory}
    \end{subfigure}
    \caption{
        \textbf{OSWorld task categories.} End-to-end latency and memory footprint for OSWorld tasks grouped by task type.
    }
    \label{fig:osworld-category}
\end{figure*}



Figure~\ref{fig:e2e-results} shows end-to-end task-latency distributions for the two CUA workloads.
Across both, \sys{} shifts the latency CDF left of CRIU and KVM throughout the distribution,
not only at the median.
The AgentLoop-on-GTA results in Figure~\ref{fig:e2e-agentloop} isolate branching cost most
clearly because many tasks are short,
so baseline overhead is a large fraction of each task:
KVM is worst because every branch requires VM-level cloning,
CRIU improves on KVM but still pays synchronous checkpoint/restore on the critical path,
and \sys{} removes this on-path cost by making branches
runnable immediately and deferring durable checkpointing.

The Agent-S3 on OSWorld results in Figure~\ref{fig:e2e-osworld} cover longer,
more diverse desktop tasks with GUI state, browser profiles, application files,
and multi-step trajectories. Branching overhead remains visible in end-to-end runtime even here,
and \sys{} completes tasks earlier with shorter tails than KVM and CRIU.
The gains are largest in this setting because the desktop workspace is heavier:
\sys{} is up to 1.9$\times$ faster than KVM and 1.5$\times$ faster than CRIU; the biggest improvements fall on
short tasks, where substrate overhead is the dominant fraction of runtime.
Figure~\ref{fig:osworld-category} breaks OSWorld down by category and reports both latency
and memory footprint.
Browser and office tasks carry larger application and filesystem state,
and multi-app tasks span several active applications with larger cross-application state;
these amplify clone and restore cost. \sys{} reduces latency across all categories,
with the largest gains on multi-app tasks, because unchanged anonymous memory and
read-mostly filesystem state are shared across branches rather than materialized independently.
Memory footprint follows the same trend: KVM is highest because each branch carries VM-level state,
CRIU materializes more branch-local state during checkpoint/restore,
and \sys{} stays lowest by preserving shared state across branches.




Overall, these results show that workspace branching must remain lightweight on the execution path. Systems based on synchronous checkpoint/restore or full VM cloning preserve execution state, but their branch overhead accumulates under rollback and speculative execution. \sys{} reduces this overhead sufficiently for versioned workspaces to remain practical during normal CUA execution.

\subsection{Microbenchmarks}

{
\begin{figure*}[th]
\begin{minipage}{0.32\textwidth}
\begin{center}
\centerline{\includegraphics[width=\linewidth]{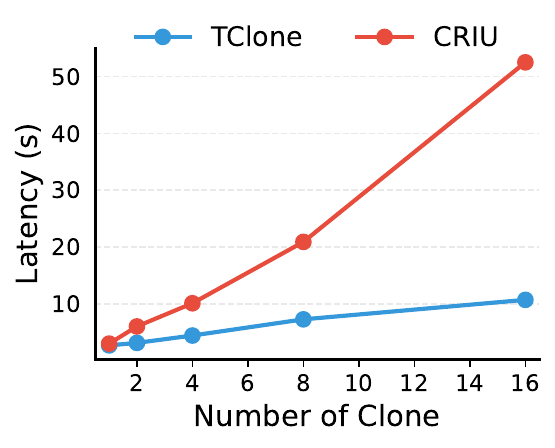}}
\vspace{0.1in}
\caption{Clone latency vs. number of concurrent clone}
\label{fig:scalability-latency}
\end{center}
\end{minipage}
\hfill
\begin{minipage}{0.32\textwidth}
\begin{center}
\centerline{\includegraphics[width=\linewidth]{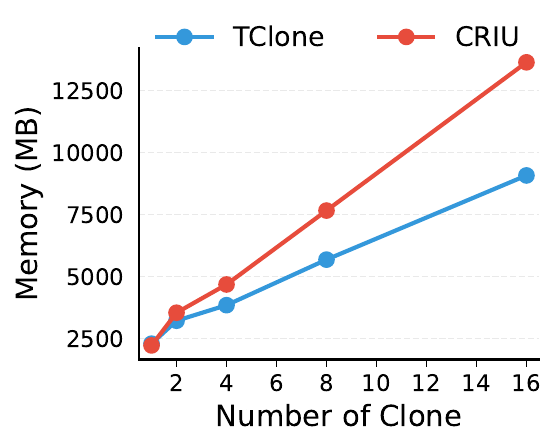}}
\vspace{0.1in}
\caption{Memory footprint vs. number of concurrent clone}
\label{fig:scalability-memory}
\end{center}
\end{minipage}
\hfill
\begin{minipage}{0.32\textwidth}
\begin{center}
\centerline{\includegraphics[width=\linewidth]{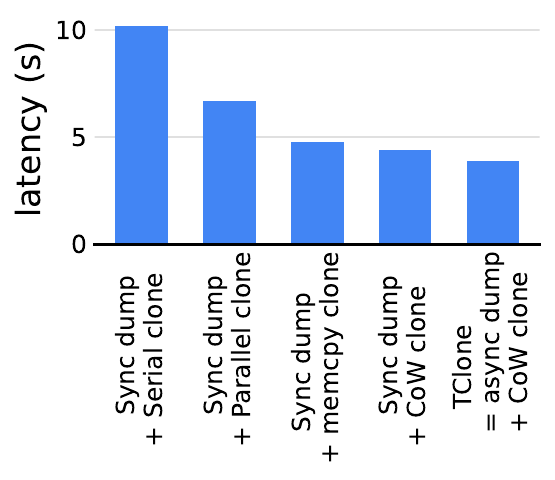}}
\caption{Performance Breakdown of \sys\ running 4 clones}
\label{fig:ablation-latency}
\end{center}
\end{minipage}
\end{figure*}
}
\begin{figure}[t]
    \centering
    \includegraphics[width=\linewidth]{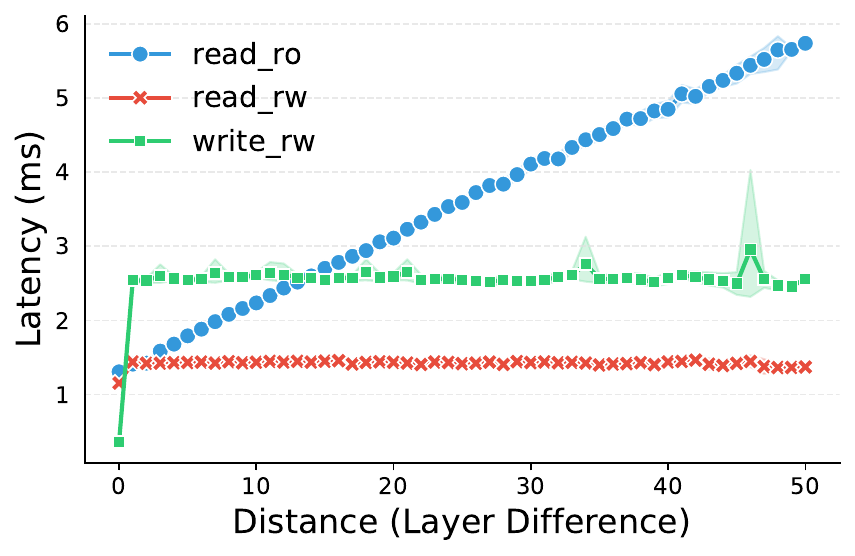}
    \caption{File Operation Latency vs. OverlayFS Layers.} 
    \label{fig:layer-distance}
\end{figure}

\subsubsection{Workspace Clone Scalability}
Figure~\ref{fig:scalability-latency} isolates substrate cost as the runtime keeps
more concurrent workspace clones alive, the regime that matters for best-of-\(N\),
beam-style execution, and rollback-heavy search. \sys{} scales substantially better than CRIU:
at 16 concurrent clones CRIU exceeds 50\,s while \sys{} stays near 10\,s, roughly a 5$\times$ gap
that widens with clone count. CRIU repeats a synchronous checkpoint/restore for every clone,
while \sys{} creates each branch through shared state and defers durable checkpointing off the fast path,
so per-branch cost grows with the divergent working set rather than with total workspace size.
The residual per-branch cost at this scale is dominated by CRIU's
per-process metadata collection: CRIU reconstructs each process by parsing its
\texttt{procfs} entries (memory maps, file descriptors, and status), whose cost scales
with the number of processes managed in total including both source workspace and cloned workspaces.

Figure~\ref{fig:scalability-memory} shows the corresponding footprint.
Both systems grow with clone count, but \sys{} stays consistently lower (about 9\,GB
versus roughly 14\,GB for CRIU at 16 clones) because \sys{} shares unchanged anonymous and
read-mostly file-backed state copy-on-write while CRIU re-materializes the full workspace
per clone. \sys{}'s growth comes only from genuinely non-shareable per-branch state: branch-local
GUI framebuffers, the memory and processes the agent newly creates as it acts, and per-branch
kernel and namespace bookkeeping. Footprint therefore tracks divergence rather than total
workspace size, so \sys{} sustains wider speculative search before memory becomes prohibitive.

\subsubsection{Ablation Study}
\label{sec:eval-ablation}

Figure~\ref{fig:ablation-latency} decomposes \sys{}'s clone latency for a Chromium workspace
with 168 running processes, creating a beam of four branches; without sharing,
each branch would duplicate roughly 2\,GB of memory.
Each bar enables one mechanism from \S\ref{sec:design},
moving from a stock CRIU path toward \sys{}.
The leftmost bar is stock CRIU: a synchronous dump followed by serial restores.
The second restores branches in parallel,
removing CRIU's restriction that only one restore proceeds while the dump state is locked.
The third overlaps the synchronous dump with the restore-side memory copy so the two proceed together,
though the clone still waits on the slower of them. The fourth replaces memory copies with copy-on-write
page sharing, cutting memory-copy cost but still blocking on the synchronous disk dump.
The final bar is \sys{}: durable dumping is moved off the branch-creation path and unchanged memory 
is shared copy-on-write. The progression shows that parallelism and overlap help,
but the decisive reductions come from the two design choices the ablation isolates:
making a branch runnable before durable serialization completes,
and sharing the large memory footprint rather than copying it.



\subsubsection{File System Overhead}
Section~\ref{sec:design-fs} argued that OverlayFS is unsuitable for CUA workloads
because read lookup latency degrades with layer depth.
CUA branching is read-dominated, so per-branch cost is set by read latency as
branch chains deepen. Figure~\ref{fig:layer-distance} measures
per-operation latency on a 1\,MB contiguous file across branch depth 0 to 50. Reads and
writes of a page already materialized branch-local (\texttt{read\_rw}, \texttt{write\_rw})
are flat across depth (roughly 1.4\,ms and 2.5\,ms), since they do not traverse ancestors.
Only a cold shared read (\texttt{read\_ro}) walks the chain, growing linearly with no
per-layer compounding from about 1.3\,ms to about 5.7\,ms at depth 50. Thus,
OverlayFS inflates every lower-layer read as the union deepens, unsuitable for read-heavy and snapshot frequent CUA branching.

\section{Related Work}
\label{sec:related}

\subsection{Agent Execution Environments}

Recent agent systems increasingly rely on managed execution environments to run commands, manipulate files, install dependencies, and interact with external tools. Software-engineering agents such as OpenAI Codex and SWE-agent execute code inside sandboxed or containerized workspaces, while OpenAI Sandbox Agents provide resumable environments with files, packages, commands, ports, and snapshots~\cite{openaicodexsandbox,openaiagentsandbox,sweagent}. Anthropic's computer-use reference environment similarly runs the agent loop inside a controlled virtual desktop environment~\cite{claudecomputeruse}. These systems show that execution substrates are central to agent safety and usability.

However, existing agent execution environments are typically designed as clean sandboxes or task-specific workspaces. They provide isolation from the host and some support for persistence, but they do not expose versioning as an online execution primitive. In particular, they do not support low-latency forking of a live personal workspace, copy-on-write sharing across speculative branches, rollback to intermediate workspace versions, or controlled merge with a concurrent human branch. \sys{} targets this missing execution substrate: it treats the workspace itself as a versioned runtime object that agents can fork, search over, roll back, and commit.

\subsection{VM-Based Snapshots and Migration}

Virtual machines provide a natural unit for capturing rich execution state. Commercial VM snapshot mechanisms can preserve virtual disk state and, optionally, memory and device state, while live-migration systems such as VMware's fast transparent migration and Xen live migration move running VMs across machines with low downtime~\cite{vmwaresnapshots,nelson2005fast,clark2005live}. Remus extends this model by asynchronously replicating VM state to provide high availability~\cite{cully2008remus}. These systems are designed primarily for migration, backup, and fault tolerance: they preserve or move a running machine, but they do not make frequent branching a first-class online operation.

Other VM systems support faster cloning. Potemkin uses flash cloning to create many virtual honeypots, SnowFlock introduces VM fork for rapid cloud cloning, and Parallax provides efficient virtual disk snapshots~\cite{vrable2005potemkin,lagarcavilla2009snowflock,meyer2008parallax}. These systems are closest in spirit to \sys{} because they exploit sharing between related VM instances. However, they target VM-level workloads such as cloud workers, honeypots, or virtual disks. \sys{} instead targets live personal workspaces used by computer-use agents: GUI sessions, browser state, credentials, application memory, filesystem state, and concurrent human activity must be versioned together. Its goal is not only to clone or migrate a machine, but to support agent-native operations such as fork, rollback, select, merge, and commit over interactive workspace branches.

\subsection{Container and LightVM Checkpoint}

Container checkpoint/restore systems preserve process state at a finer granularity than full VMs. CRIU can freeze Linux applications or containers and restore process trees, memory, namespaces, file descriptors, and related kernel state; Docker checkpoint exposes this functionality for container workloads~\cite{criu,dockercheckpoint}. \sys{} follows the same general reconstruction style when rebuilding a process tree inside a new workspace container. However, CRIU-style checkpoint/restore treats cloning as a synchronous save-and-restore operation, which is suitable for migration or recovery but too expensive for speculative agent execution. \sys{} separates fast branch creation from durable state capture: it creates runnable branches using copy-on-write sharing and performs checkpointing asynchronously.

Lightweight VM and sandboxing systems such as Firecracker, Kata Containers, and gVisor improve the provide stronger isolation than regular containers and lower overhead than traditional VMs~\cite{firecracker,kata,gvisor}. However, they still expose execution environments as isolated instances rather than as versioned personal workspaces. They do not directly provide cross-branch copy-on-write memory sharing, branch-aware filesystem state, GUI-local checkpointing, rollback to intermediate agent states, or merge with a concurrent human branch. \sys{} is complementary to these substrates: it focuses on the workspace-versioning semantics required for computer-use agents, while lower-level container or LightVM mechanisms can serve as isolation building blocks.
\section{Conclusion}

Computer-use agents need to operate over rich personal workspaces without inheriting all of the risk of direct execution on a user's everyday PC. \sys{} introduces a versioned workspace substrate that can fork a live GUI workspace into isolated branches, roll back failed trajectories, and promote selected results. By separating fast branch creation from durable checkpointing and using copy-on-write sharing for memory and filesystem state, \sys{} makes rollback and speculative execution practical for interactive agent workloads. Our evaluation on OSWorld and AgentLoop workloads shows that these mechanisms reduce end-to-end task latency compared with CRIU-based checkpoint/restore and KVM-based cloning.

\bibliographystyle{plain}
\bibliography{sysml,all-defs,references}

\end{document}